\documentclass[12pt,aps,prd,preprint,tightenlines,superscriptaddress,
amsmath,amssymb,nofootinbib]{revtex4}

\RequirePackage[colorlinks=true
,urlcolor=blue
,anchorcolor=blue
,citecolor=blue
,filecolor=blue
,linkcolor=blue
,menucolor=blue
,linktocpage=true
,pdfproducer=medialab
,pdfa=true
]{hyperref}

\allowdisplaybreaks

\usepackage{amsmath,amssymb,amsthm,amsfonts}
\usepackage{graphicx}
\usepackage{subfigure}
\usepackage{dcolumn}	
\usepackage{hyperref}      	
\usepackage{bm}		
\usepackage{epsfig}
\usepackage{epstopdf}
\usepackage{setspace}
\usepackage[usenames, dvipsnames]{color}
\usepackage{slashed}
\usepackage{comment}
\usepackage{enumitem}

\newcommand{\PRE}[1]{{#1}} 

\newcommand{\be}{\begin{equation}\begin{aligned}}
\newcommand{\ee}{\end{aligned}\end{equation}}
\newcommand{\beq}{\begin{equation}}
\newcommand{\eeq}{\end{equation}}
\newcommand{\beqa}{\begin{eqnarray}}
\newcommand{\eeqa}{\end{eqnarray}}
\newcommand{\nn}{\nonumber}

\newcommand{\ifb}{\text{fb}^{-1}}
\newcommand{\iab}{\text{ab}^{-1}}

\newcommand{\kev}{\text{keV}}
\newcommand{\mev}{\text{MeV}}
\newcommand{\gev}{\text{GeV}}
\newcommand{\tev}{\text{TeV}}

\newcommand{\micm}{\mu\text{m}}
\newcommand{\mm}{\text{mm}}
\newcommand{\cm}{\text{cm}}
\newcommand{\m}{\text{m}}

\newcommand{\mrad}{\text{mrad}}
\newcommand{\murad}{\mu\text{rad}}

\newcommand{\eg}{{\em e.g.}}
\newcommand{\ie}{{\em i.e.}}

\renewcommand{\eqref}[1]{Eq.~(\ref{#1})}

\newcommand{\secref}[1]{Sec.~\ref{sec:#1}}

\newcommand{\figref}[1]{Fig.~\ref{fig:#1}}

\newcommand{\appref}[1]{Appendix~\ref{sec:#1}}

\newcommand{\TAN}{TAXN}
\newcommand{\lmin}{L_{\text{min}}}
\newcommand{\lmax}{L_{\text{max}}}
\newcommand{\ltan}{L_{\text{\TAN}}}
\newcommand{\aang}{{\cal A}_{\text{ang}}}

\newcommand{\gpi}{g_{\pi^0\gamma\gamma}}
\newcommand{\geta}{g_{\eta\gamma\gamma}}
\newcommand{\ga}{g_{a\gamma\gamma}}

\begin{document}

\preprint{UCI-TR-2018-02}

\title{\PRE{\vspace*{1.0in}}
ALPs at FASER: The LHC as a Photon Beam Dump
\PRE{\vspace*{.4in}}}

\author{Jonathan L.~Feng}
\email{jlf@uci.edu}
\affiliation{Department of Physics and Astronomy, University of
California, Irvine, CA 92697-4575 USA
\PRE{\vspace*{.1in}}}

\author{Iftah Galon}
\email{iftah.galon@physics.rutgers.edu}
\affiliation{New High Energy Theory Center \\
Rutgers, The State University of New Jersey \\
Piscataway, New Jersey 08854-8019, USA
\PRE{\vspace*{.1in}}}

\author{Felix Kling}
\email{fkling@uci.edu}
\affiliation{Department of Physics and Astronomy, University of
California, Irvine, CA 92697-4575 USA
\PRE{\vspace*{.1in}}}

\author{Sebastian Trojanowski\PRE{\vspace*{.2in}}}
\email{Sebastian.Trojanowski@ncbj.gov.pl}
\affiliation{Department of Physics and Astronomy, University of
California, Irvine, CA 92697-4575 USA
\PRE{\vspace*{.1in}}}
\affiliation{National Centre for Nuclear Research,\\Ho{\. z}a 69, 00-681 Warsaw, Poland
\PRE{\vspace*{.4in}}}


\begin{abstract}
\PRE{\vspace*{.2in}}
The goal of FASER, ForwArd Search ExpeRiment at the LHC, is to discover light, weakly-interacting particles with a small and inexpensive detector placed in the far-forward region of ATLAS or CMS.  A promising location in an unused service tunnel 480 m downstream of the ATLAS interaction point (IP) has been identified.  Previous studies have found that FASER has significant discovery potential for new particles produced at the IP, including dark photons, dark Higgs bosons, and heavy neutral leptons.  In this study, we explore a qualitatively different, ``beam dump'' capability of FASER, in which the new particles are produced not at the IP, but through collisions in detector elements further downstream.  In particular, we consider the discovery prospects for axion-like particles (ALPs) that couple to the standard model through the $a \gamma \gamma$ interaction.  TeV-scale photons produced at the IP collide with the \TAN\ neutral particle absorber 130 m downstream, producing ALPs through the Primakoff process, and the ALPs then decay to two photons in FASER.   We show that FASER can discover ALPs with masses $m_a \sim 30 - 400~\mev$ and couplings $\ga \sim 10^{-6} - 10^{-3}~\gev^{-1}$, and we discuss the ALP signal characteristics and detector requirements.
\end{abstract}


\maketitle


\clearpage

\section{Introduction}
\label{sec:introduction}

The Large Hadron Collider (LHC) has already played a remarkable role in pushing back the boundaries of our knowledge about particle physics. At the same time, no evidence for physics beyond the standard model (SM) has yet emerged.  For decades, the focus has been on new physics at the weak scale, given significant motivations from the gauge hierarchy problem, weakly interacting massive particle (WIMP) dark matter, and low-energy anomalies.  More recently, however, there has been a growing, complementary interest in new particles that are light and weakly-coupled~\cite{Battaglieri:2017aum}.  Like WIMPs, such particles may have the correct thermal relic density~\cite{Boehm:2003hm}, and in fact, light particles with MeV to GeV masses and the correct thermal relic density emerge naturally in theories motivated by the gauge hierarchy problem~\cite{Feng:2008ya}.  In addition, such particles may resolve outstanding anomalies in low-energy data~\cite{Bennett:2006fi,Marciano:2016yhf,Pohl:2010zza,Krasznahorkay:2015iga}.  Perhaps most importantly, new, light particles can be searched for with relatively inexpensive, small, and fast experiments, opening the floodgates to a host of experimental proposals with potentially revolutionary implications for particle physics and cosmology.

If new particles are light and weakly-interacting, they may be copiously produced in proton-proton collisions, but they preferentially go in the forward direction and escape detection at the ATLAS~\cite{Aad:2008zzm} and CMS~\cite{Chatrchyan:2008aa} experiments. This suggests that novel experimental proposals may be able to enhance the LHC's discovery potential.  In recent papers~\cite{Feng:2017uoz,Feng:2017vli,Kling:2018wct}, we have proposed and explored the potential of one such experiment: FASER, ForwArd Search ExpeRiment at the LHC.  These studies and others~\cite{Batell:2017kty,Helo:2018qej,Bauer:2018onh} have established FASER's significant potential to discover a host of proposed particles, including dark photons and other light gauge bosons, dark Higgs bosons and other scalar mediators, and sterile neutrinos or heavy neutral leptons (HNLs). 

FASER is a small detector, with volume $\sim 1~\m^3$, that will be placed along the beam collision axis or line of sight (LOS), hundreds of meters downstream from the ATLAS or CMS interaction point (IP).  A particularly promising location is a few meters to the side of the main LHC tunnel, 480 m downstream from the ATLAS IP in service tunnel TI18.  This tunnel was formerly used to connect the SPS to the LEP tunnel, but is currently empty and unused. In this location, FASER exploits the enormous, previously ``wasted'' cross section for very forward physics ($\sigma \sim \text{100 mb}$), which implies that even very weakly-coupled new particles can be produced in large numbers at the LHC.  In addition, the production of long-lived particles (LLPs) at high center-of-mass energy results in high boosts, leading to long propagation distances ($\bar{d} \sim  {\cal O}(100)~\m$) and decays that are far beyond the main LHC infrastructure and in places where backgrounds are expected to be highly suppressed.  In addition to FASER, there are a number of other current and proposed experiments that target particles with similar properties, including NA62~\cite{NA62:2017rwk}, SeaQuest~\cite{Gardner:2015wea,Berlin:2018pwi}, SHiP~\cite{Alekhin:2015byh}, MATHUSLA~\cite{Chou:2016lxi,Curtin:2017izq}, and CODEX-b~\cite{Gligorov:2017nwh}.

In our previous work, we studied FASER's potential to detect dark photons~\cite{Feng:2017uoz}, dark Higgs bosons~\cite{Feng:2017vli}, and HNLs~\cite{Kling:2018wct}.  In all of these cases, the new physics couples to the SM through dimension-4 interactions (renormalizable portals), and the new particles are produced at the IP.  In this work, we consider a qualitatively different possibility: axion-like particles (ALPs), which couple to the SM through dimension-5 interactions and are created not at the IP, but further downstream.  ALPs are pseudoscalar SM-singlets.  As with the QCD axion~\cite{Peccei:1977hh,Peccei:1977ur,Wilczek:1977pj,Weinberg:1977ma}, they can appear as pseudo-Nambu-Goldstone bosons in theories with broken global symmetries. (For a recent review, see, \eg, Ref.~\cite{Bauer:2017ris}.)   When the global symmetry is broken at a high energy scale $f$, highly suppressed dimension-5 interactions are generated between ALPs and SM gauge bosons, $(1/f)\,a V^{\mu\nu} \widetilde V_{\mu\nu}$~\cite{Georgi:1986df}.  As in the case of the QCD axion, a shift symmetry $a \to a + c$ can naturally keep the ALP mass low.  On the other hand, for generalized ALPs, one typically introduces a small mass term $\frac{1}{2} m_a^2 a^2$ that softly breaks the shift symmetry and allows $m_a$ to be an independent parameter of the model. Low-mass ALPs with suppressed couplings to the SM are then long-lived particles (LLPs) that can be sensitively probed by FASER.

\begin{figure}[t]
\centering
\includegraphics[width=0.99\textwidth]{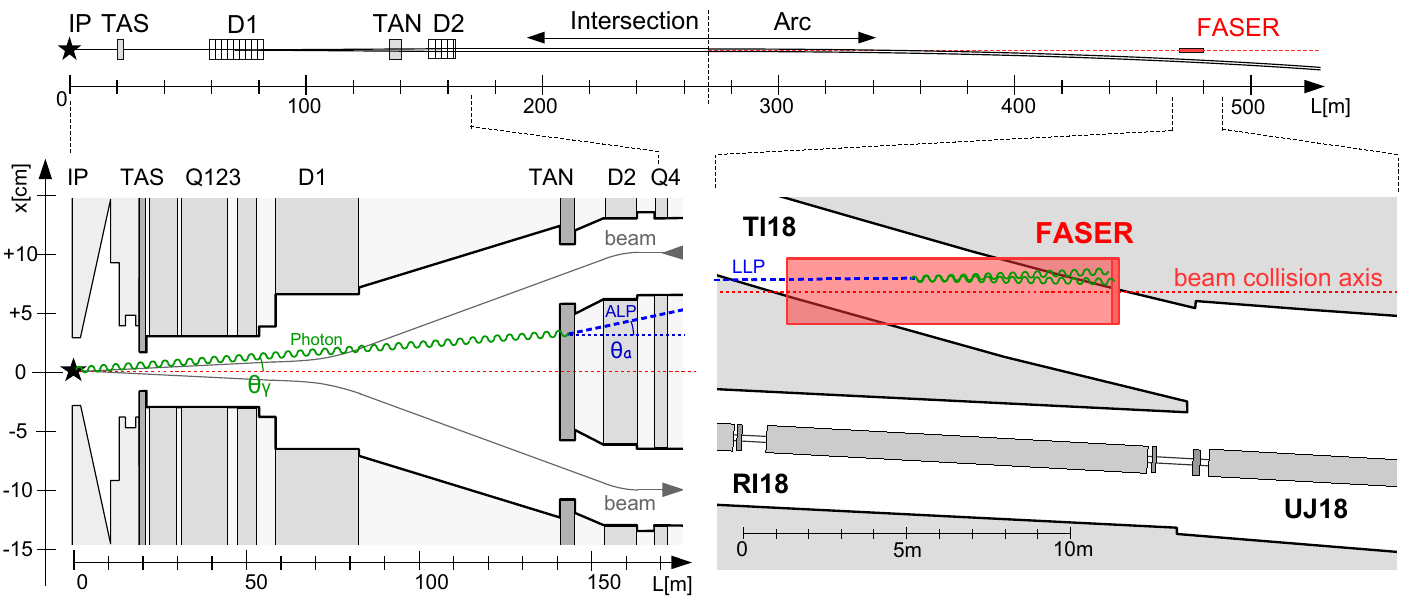} 
\caption{The dominant ALP production and decay processes considered in this study.  {\bf Upper panel}: FASER is located $480~\m$ downstream from the IP along the beam collision axis (dotted line) after the main LHC tunnel curves away.  {\bf Lower left panel}: High-energy photons are produced at the IP with small angles $\theta_{\gamma}$ relative to the beamline.  These photons then collide with the neutral particle absorber (TAN or \TAN), producing ALPs $a$ at similarly small angles $\theta_a$ relative to the beamline.  Note the extreme difference in horizontal and vertical scales.  {\bf Lower right panel}: The ALPs then travel $\sim 350~\m$ further downstream and decay through $a \to \gamma \gamma$ to two highly collinear, high-energy photons in FASER, which is located in the side tunnel TI18 close to the UJ18 hall. }
\label{fig:setup}
\end{figure}

An interesting possibility, which leads to a phenomenology that is qualitatively different from the models considered in our previous studies, is that ALPs are predominantly coupled to two photons~\cite{Dobrich:2015jyk}. In this case, ALPs can be produced in $pp$ collisions at the IP through, \eg, photon fusion or rare decays of neutral pions. However, as we show below, for high-energy forward-going ALPs that can reach FASER, the dominant production process is one in which photons produced at the IP collide with elements of the LHC infrastructure $\sim 130~\m$ downstream, producing ALPs through the Primakoff process $\gamma N \to a N' X$~\cite{Primakoff:1951pj,Tsai:1986tx}.  The ALPs then travel another $\sim 350~\m$ and decay to two photons in FASER.  This process, through which the LHC can be thought of as a {\it high-energy photon beam dump experiment}, is depicted in \figref{setup}. 

In this study, we evaluate the prospects for FASER to discover ALPs that are produced through their di-photon coupling and decay through $a \to \gamma \gamma$.  This work is structured as follows. In \secref{properties} we review the basic properties of ALPs and their di-photon coupling. In \secref{productionadndecay} we discuss ALP production and decay. The discovery reach of FASER for ALPs is presented in \secref{results}.  In \secref{detection}, we discuss the detector requirements for detecting the ALP signal in FASER. Our conclusions are collected in \secref{conclusions}.  Details of the kinematics of the Primakoff process, ALP production in rare meson decays, and the angular acceptance function for FASER are given in Appendices \ref{sec:Primakoffkinematics}, \ref{sec:piondecay}, and \ref{sec:angularacceptance}, respectively.

\section{Properties of Axion-Like Particles}
\label{sec:properties}

We consider a low-energy effective theory in which an ALP $a$ couples to vector bosons through the dimension-5 interactions
\be
{\cal L}_{\text int} = -\frac14 g_{aBB}\, a B_{\mu\nu}\widetilde B^{\mu\nu} 
-\frac14 g_{aWW}\, a W^A_{\mu\nu}\widetilde W^{A, \, \mu\nu} \ ,
\ee
where $B_{\mu\nu}$ and $W^A_{\mu\nu}$ are the U(1)$_Y$ and SU(2)$_L$ field strength tensors, respectively, and $g_{aBB}$ and $g_{aWW}$ are the corresponding coupling constants with dimension $\gev^{-1}$. If such interactions are generated by physics with coupling $\alpha$ that was integrated out at some heavy scale $f$, one expects
\be
g_{aBB}, \ g_{aWW} \sim \frac{\alpha}{2\pi f} \ .
\ee
This is the case for axions~\cite{Peccei:1977hh, Peccei:1977ur, Wilczek:1977pj, Weinberg:1977ma}, for example, and more generally for pseudo-Goldstone bosons with non-vanishing axial anomalies.

After electroweak symmetry breaking, the couplings $g_{aBB}$ and $g_{aWW}$ induce couplings of the ALP to $\gamma\gamma$, $\gamma Z$, $ZZ$, and $W^+W^-$. In this study, we focus on the $\gamma \gamma$ coupling and neglect the others. The other couplings typically have a small effect on our signal; for example, the subdominant production process of ALPs from meson decays can be enhanced by $W^+W^-$ couplings~\cite{Izaguirre:2016dfi}. More important are their effects on other observables.  For example, a non-vanishing $\gamma Z$ coupling induces the exotic decay $Z\to a\gamma$. Although this process does not contribute significantly to the ALP production rate in the far forward region, it can be searched for in high-$p_T$ experiments at the LHC. (See Refs.~\cite{Bauer:2017ris,Jaeckel:2015jla} for some future projections.) 

ALPs may also couple through dimension-5 operators to gluons and fermions, as well as through dimension-6 couplings to the Higgs boson. For a recent review see, \eg, Ref.~\cite{Bauer:2017ris}. In this study, we will assume that the effects of these other couplings on our signal processes are negligible.  This is the case when these couplings are relatively small, or, for example, when the couplings are to heavy particles, such as third-generation fermions, and so their impact on ALP production and decay is suppressed.  It is important to note, however, that gluon and fermion couplings generate di-photon couplings through loops and vice versa, and so to analyze a specific underlying ALP model in detail, one would in general have to include all of these couplings in a unified way.  Here, we take a more model-independent, phenomenological approach.

With these simplifying assumptions, we therefore focus on the ALP effective Lagrangian
\be
\mathcal{L} \supset \frac{1}{2} \partial_\mu a \, \partial^\mu a - \frac{1}{2} m_a^2 a^2 
- \frac{1}{4} \ga a F^{\mu\nu} \widetilde F_{\mu\nu} \ ,
\label{eq:LafterEWSB}
\ee
where $F_{\mu\nu}$ is the field strength tensor of electromagnetism.  The resulting parameter space is very simple, as it is spanned by two parameters: the ALP mass $m_a$ and the di-photon coupling $\ga$. 

With this Lagrangian, the ALP decay width is
\be
\Gamma_a \equiv \Gamma(a\to\gamma\gamma) = \frac{\ga^2 m_a^3}{64\pi} \ .
\ee
The cubic dependence on $m_a$, resulting from the fact that the decay is mediated by a dimension-5 operator, implies that, as the ALP mass decreases, the ALP lifetime increases rapidly.  The ALP decay length is
\be
\bar{d}_a= \frac{c}{\Gamma_a} \gamma_a\beta_a
\approx 630~\m \,
\bigg[\frac{10^{-4}~\gev^{-1}}{\ga}\bigg]^2 
\bigg[\frac{p_a}{\tev}\bigg] 
\bigg[\frac{50~\mev}{m_a}\bigg]^4 \ ,
\label{eq:dbar}
\ee
where we have normalized to currently viable values of $\ga$ and $m_a$.  For these values and ALP momenta $p_a\sim\tev$, the ALP decay length is naturally hundreds of meters, \ie, in the range relevant for FASER searches for LLPs. 

Although the ALP decays primarily into pairs of photons, it is possible that one of the photons converts into an electron pair leading to the decay $a \to e^+ e^- \gamma$.  The branching fraction for this decay is~\cite{Terschlusen:2013iqa}
\be
\mathcal{B}(a \to e^+e^- \gamma) = \frac{e^2}{6\pi} \int_{4m_e^2}^{m_a^2} \frac{dq^2}{q^2}  |F(q^2)|^2
\bigg[1-\frac{4\m_e^2}{q^2} \bigg]^{\frac{1}{2}}
\bigg[1+\frac{2\m_e^2}{q^2} \bigg]
\bigg[1-\frac{q^2}{m_a^2} \bigg]^{3} \ ,
\ee 
where $q^2$ is the invariant mass of the electron pair, and $F(q^2)\approx 1$. For ALP masses between $m_a = 10~\mev $ and $ 1~\gev$, the branching fraction ranges from $\mathcal{B}(a\to e^+e^-\gamma) = 0.4\%$ to  $1.7\%$. Note that this branching fraction peaks at low $q^2$, implying that most of the ALP energy will be carried by the photon, while the electrons will typically be softer.

\section{ALP Production and Decay in FASER}
\label{sec:productionadndecay}

\subsection{Mechanisms for ALP Production in the Forward Region}

\begin{figure}[t]
\centering
\includegraphics[width=0.32\textwidth]{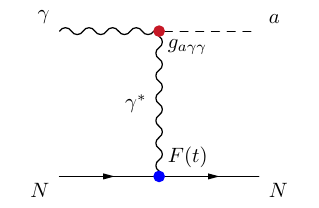} 
\hspace*{2cm}
\includegraphics[width=0.32\textwidth]{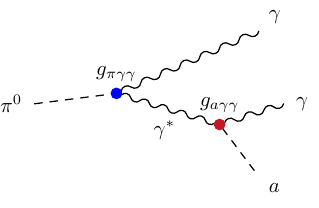} 
\caption{ALP production in the Primakoff process (left) and light meson decay (right).}
\label{fig:alp_production}
\end{figure}

In the dominant ALP-photon coupling scenario, ALPs can be produced in any process involving photons by radiating an ALP off a photon line. However, for FASER, we are primarily interested in the production of highly energetic ALPs in the very forward region. The dominant production mechanism is then the Primakoff process, in which a photon converts into an ALP when colliding with a nucleus. This can happen when photons produced at the LHC collide with the forward LHC infrastructure, as illustrated in \figref{setup}. The corresponding Feynman diagram is shown in the left panel of \figref{alp_production}.

Photons are produced at the IP mainly through $\pi^0$ decay.  They then propagate in the beam pipe until they hit the material of the LHC accelerator, as illustrated in the bottom left panel of \figref{setup}. Very forward photons collide with the neutral particle absorber, which is designed to protect the magnets behind it.  In the current LHC, this is the TAN, a $\sim 3.5~\m$ thick metal block placed along the beam collision axis at a distance of $140~\m$ from the IP.  At the high-luminosity LHC (HL-LHC) this absorber is planned to be upgraded to the \TAN\ and shifted to a new position about $130~\m$ away from the IP~\cite{Apollinari:2116337}. In the following we will use the details of the upgraded absorber \TAN.  The reach of FASER is only mildly sensitive to the precise properties and location of the absorber. 

Very forward ALPs may also arise from the exotic decays of light mesons, shown in the right panel of \figref{alp_production}, which are abundantly produced at the IP with very high forward-going momenta. However, as we discuss in more detail in \appref{piondecay}, such rare meson decays typically give a subdominant contribution to the FASER signal relative to the Primakoff process.  In the rest of this section we therefore focus primarily on the Primakoff process.  When presenting FASER's sensitivity reach in \secref{results}, however, we include the dominant exotic meson decays, $\pi^0 \to a\gamma\gamma$ and $\eta \to a\gamma\gamma$.

Last, we note that ALPs may also be produced at the LHC through other processes, \eg, through  exotic $Z$ decays or photon fusion.\footnote{An interesting approach is to use ATLAS and CMS to trigger on ultra-peripheral heavy-ion collisions, which makes it possible to search for ALPs produced in photon fusion~\cite{Knapen:2016moh,Baldenegro:2018hng}.} As with rare meson decays, however, these processes do not typically produce large numbers of boosted forward-going ALPs, and they are therefore subdominant contributions to the FASER signal.

\subsection{Primakoff Process in the \TAN}

\begin{figure}[t]
\centering
\includegraphics[width=0.32\textwidth]{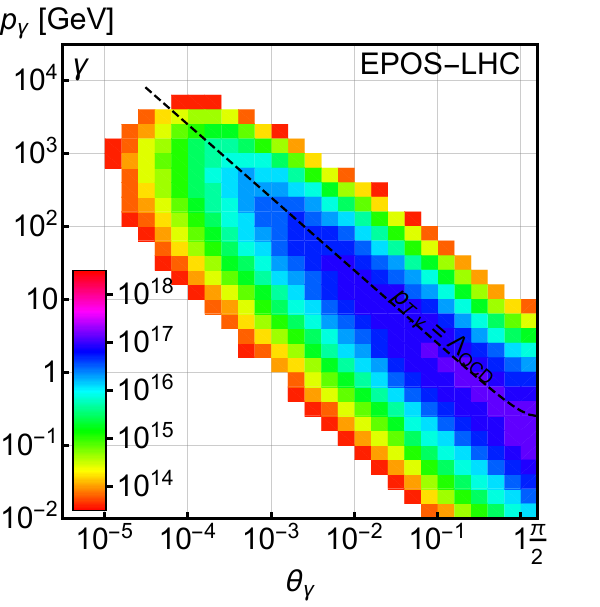}
\includegraphics[width=0.32\textwidth]{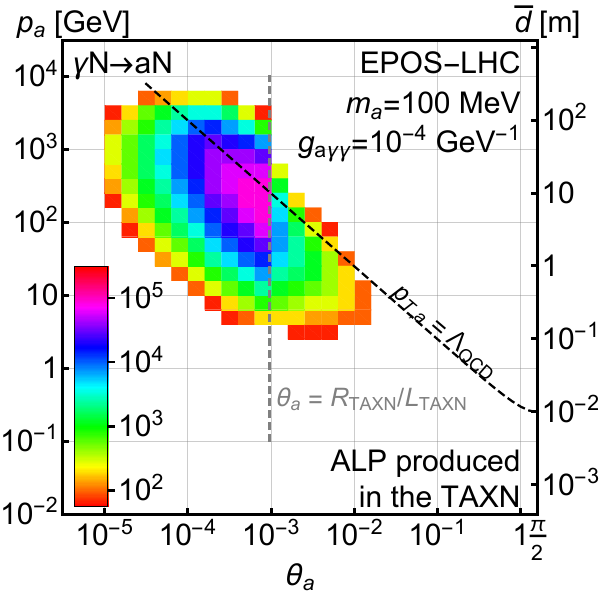}
\includegraphics[width=0.32\textwidth]{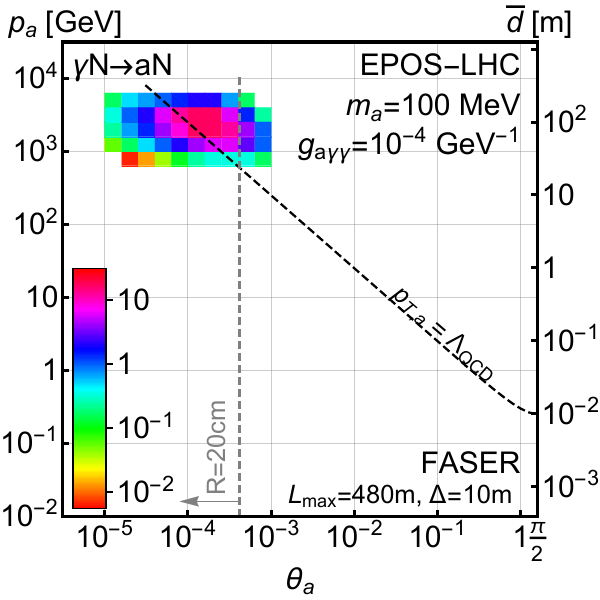}
\caption{Distributions of photons (left), ALPs produced in the \TAN\ (center), and ALPs decaying at FASER's position within the range $(\lmin, \lmax)$ (right) in the $(\theta, p)$ plane, where $\theta$ is the particle's angle with respect to the beam axis, and $p$ is the particle's momentum.  We assume the HL-LHC integrated luminosity $3~\iab$ and a \TAN\ radius of $R_{\text{\TAN}} = 12.5~\cm$, which implies that ALPs produced at the \TAN\ typically have $\theta_a < R_{\text{\TAN}} / L_{\text{\TAN}} \approx 1~\mrad$. }
\label{fig:PvsT}
\end{figure}

Forward high-energy photons that eventually hit the \TAN\ are copiously produced in $pp$ collisions at the IP, primarily in meson decays. To estimate the FASER event yield of ALPs produced by such photons, a reliable estimate of the forward photon spectrum is required. In the left panel of \figref{PvsT} we show the estimated photon spectrum in the  $(\theta_\gamma,p_\gamma)$ plane, where $\theta_\gamma$ and $p_\gamma$ are the photon's angle with respect to the beam axis and its momentum, respectively. The spectrum was simulated using the CRMC package~\cite{CRMC}, applying the EPOS-LHC model~\cite{Pierog:2013ria}. This includes photons produced in the decays of all light mesons. The dominant contribution comes from the decays $\pi^0, \eta \to\gamma\gamma$; decays of heavier mesons provide only a small correction. As can be seen, in the log-log plot the events cluster around the line defined by $p_\gamma\, \theta_\gamma \approx p_T = \Lambda_{QCD} \approx 0.25~\gev$. This is indicative of the characteristic momentum transfer scale for light meson production. As discussed in Ref.~\cite{Feng:2017uoz}, the results are consistent with other Monte Carlo simulations, such as QGSJET-II-04~\cite{Ostapchenko:2010vb} and SIBYLL 2.3~\cite{Ahn:2009wx,Riehn:2015oba}, indicating a good understanding of the forward photon spectrum at the LHC. This is not surprising, since all three of the simulations have been tuned to the LHC data collected by the LHCf Collaboration~\cite{Adriani:2015iwv,Adriani:2017jys}. 

As noted above, we assume that the \TAN\ will be located at a distance $\ltan=130~\m$ from the IP~\cite{Apollinari:2116337}. Photons produced at the IP may collide with the \TAN\ at transverse distances up to the radius $R_{\text{\TAN}} = 12.5$ cm from the beam collision axis or LOS.  Within a transverse distance of $R_{\text{\TAN}}$ from the beamline, the \TAN\ has two holes to let the beams through.  Following Ref.~\cite{TAXNgeometry}, and as illustrated in the bottom left panel of \figref{setup}, we assume these holes are circles with radii $4.25~\cm$ and center-to-center separation $14.8~\cm$.  We take this into account when implementing the \TAN\ geometry and estimating the number of photon-ALP conversions in the \TAN. 

The differential cross section for the Primakoff process is~\cite{Tsai:1986tx}
\be
\frac{d\sigma_{\rm Prim}}{d\theta_{a\gamma}}
=2\pi\alpha Z^2 F^2(t) \frac{8 \Gamma_a}{m_a^3} \frac{p_a^4 \sin^3\theta_{a\gamma}}{t^2}
=\frac{1}{4} \ga^2\alpha Z^2 F^2(t)  \frac{p_a^4 \sin^3\theta_{a\gamma}}{t^2} \ ,
\label{eq:cross_sec_alp_prod}
\ee
where $\theta_{a\gamma}$ is the lab-frame ALP-photon opening angle, $p_a$ is the lab-frame ALP momentum, $t = -q^2 = -(p_a - p_\gamma)^2$ is the momentum exchange, and $Z$ and $F(t)$ are the target atomic number and form factor, respectively. We use the elastic form factors of the atom and the coherent one for the nucleus, and we checked that contributions from inelastic and incoherent processes can be safely neglected in our case.  Following Refs.~\cite{Tsai:1973py,Patrignani:2016xqp}, we parametrize the form factors as
\be
F(t) = 
\begin{cases}
\begin{matrix}
\frac{a^2 t}{1+a^2 t} & & t < 7.39~m_e^2 \ , \\
\frac{1}{1+t/d} & & t > 7.39~m_e^2 \ ,
\end{matrix}
\end{cases}
\label{eq:form_factors}
\ee
where $a=111\, Z_{\rm nuc}^{-1/3}/m_e$, $d= 0.164 \, A_{\rm nuc}^{-2/3}~\gev^2$, $Z_{\rm nuc}$ and $A_{\rm nuc}$ are the atomic and mass numbers for the nucleus, and $m_e=511~\kev$ is the electron mass. 

The precise value of the cross section depends on the target material. Although the main inner absorber of the \TAN\ will be made of copper, it will additionally be surrounded by steel outer shielding~\cite{TAXNcomposition}. For simplicity, we use the atomic and mass numbers for iron, $Z_{\rm nuc}=26,~A_{\rm nuc}=56$, when evaluating \eqref{eq:cross_sec_alp_prod}.  This approximation is further justified by the fact that, to a good approximation, the dependence on form factors cancels out in the ratio between the Primakoff and pair-production cross sections, which is what ultimately determines the rate of ALP production. 

The Primakoff production process competes with the other photon-matter interactions. At photon energies higher than $\sim\mev$, photon conversion to an $e^+e^-$ pair in the nuclear fields dominates over all other processes. The relevant pair-production cross section in iron for the photon energies of our interest is of the order of $\sigma_{\rm conv}\simeq 5~{\rm barn}$~\cite{Patrignani:2016xqp}. The probability of a photon to convert into an ALP, $\mathcal{P}_{\gamma\rightarrow a}$, is, then, given by 
\be
\frac{d \mathcal{P}_{\gamma\rightarrow a}}{ d\theta_{a\gamma}} 
=
\frac{1}{\sigma_{\text{conv}}}  \frac{d\sigma_{\text{Prim}}}{d\theta_{a\gamma}} \ .
\ee

To be conservative, we neglect the scatterings of secondary photons produced in the electromagnetic showers inside the \TAN, which could also produce ALPs. An accurate modeling of this contribution would require dedicated simulation tools, \eg, FLUKA~\cite{Ferrari:898301} or Geant4~\cite{Agostinelli:2002hh}, to study shower development in the \TAN, taking into account its precise geometry and composition. This is beyond the scope of the current analysis, but we note that such secondary photons will have lower energies than those produced at the IP, and they will also be less collimated along the LOS.  We therefore do not expect secondary photons to drastically improve the reach of FASER. 

Importantly, the nuclear form factor typically suppresses large momentum transfers between the projectile photon and the target nucleus; that is, the target nucleus does not recoil much. As a result, ALPs tend to carry most of the photon initial momentum and follow the direction of the incoming photon.  Consequently, the angles of the ALP and parent photon relative to the LOS are very similar, and $\theta_a \approx \theta_\gamma < R_{\text{\TAN}} / L_{\text{\TAN}} \approx 1~\mrad$. This can be seen in the central panel of \figref{PvsT}.  This implies that only a few ALPs, mostly at lower energies, are produced in processes with large enough momentum transfers to produce larger $\theta_a$. Also, photons that collide with other parts of the infrastructure besides the \TAN\ do not typically end up in FASER.  As a result, to a good approximation, the number of ALPs going towards FASER is given simply by rescaling the number of photons incident on the \TAN\ by the integrated probability of the Primakoff process to occur, $\sigma_{\text{Prim}}/\sigma_{\text{conv}}$, and their resulting momenta are determined by the collinear approximation $p_a =p_\gamma$.  A more detailed discussion of the kinematics of the Primakoff process is given in \appref{Primakoffkinematics}.

\subsection{ALP Decays in FASER}

Once produced, the ALPs decay into two photons after traveling distance $\sim \bar{d}_a$. As can be seen in \eqref{eq:dbar}, for typical ALP momenta $p_a\sim \tev$, ALP masses $m_a \sim 50~\mev$, and coupling constants $\ga \sim 10^{-4}~\gev^{-1}$, $\bar{d}$ is of the order of a few hundred meters, motivating a search for ALPs at FASER. 

Following Refs.~\cite{Feng:2017uoz, Feng:2017vli, Kling:2018wct}, we assume that FASER has a cylindrical detector volume of radius $R$, which is co-centric with the beam collision axis, and has a depth $\Delta = \lmax - \lmin$, where $\lmax$ and $\lmin$ are the distances of the far and near edges of the detector to the IP. We will show results for the following detector parameters:
\be
\text{\bf FASER:} \ \lmax=480~\m,\ \Delta = 10~\m,\ R=20~\cm \ .
\label{eq:far_location}
\ee
For the parameters of interest, the event rate is linearly proportional to $\Delta$.  For reasons explained below, reducing $\Delta$ by a factor of 2 or 3 makes very little difference to the sensitivity reach in ALP parameter space.  As mentioned above, FASER will be stationed in a side tunnel, after the curving of the main LHC tunnel containing the beam pipe.  We assume that a high granularity electromagnetic calorimeter is positioned at the back of the detector, after the tracking system, and detects photons with high efficiency (see \secref{detection}). 

The probability $\mathcal{P}_{\text{det}}$ that an ALP produced at the \TAN\ subsequently decays within FASER is 
\be
\mathcal{P}_{\text{det}}
= \left[ e^{-(\lmin-\ltan)/\bar d_a}-e^{-(\lmax-\ltan)/\bar d_a} \right] \aang(\theta_{a\gamma},~\theta_\gamma) \ ,
\label{eq:decay_in_vol}
\ee
where $\ltan=130~\m$, and the detector angular acceptance $\aang(\theta_{a\gamma},~\theta_\gamma)$ is the probability that an ALP produced by a photon at the \TAN\ has a trajectory that passes through FASER, given the scattering and photon polar angles $(\theta_{a\gamma},\theta_\gamma)$.  The scattering angle $\theta_{a\gamma}$ is defined as the ALP's angle relative to the photon direction, and the photon polar angle $\theta_\gamma$ is defined relative to the beam collision axis. In the aforementioned collinear approximation, $p_a = p_\gamma$, the angular acceptance can simply be written as $\aang(\theta_{a\gamma},~\theta_\gamma) = \Theta(\lmax \tan\theta_\gamma - R)$, where $\Theta(x)$ is the Heaviside theta function. In the special case of a cylindrical detector it is even possible to obtain an analytic solution, which is presented in \appref{angularacceptance}.  In practice, we obtain a more accurate description for the angular acceptance function through Monte Carlo simulation.

The spectrum of ALPs decaying within a distance $(\lmin,\lmax)$ from the IP is shown in the right panel of \figref{PvsT}. As we can see, only very forward ALPs with $\theta_a < R_{\text{\TAN}} / L_{\text{\TAN}}$ contribute to the signal. This allows us to have a relatively small calorimeter with radius $\sim {\cal O}(10)~\cm$, which can detect almost all available ALPs. In the following we will assume that the radius of the calorimeter is $R=20~\cm$ and coincides with the radius of FASER's decay volume.

\section{Sensitivity Reach of FASER}
\label{sec:results}

\begin{figure}[t]
\centering
\includegraphics[width=0.48\textwidth]{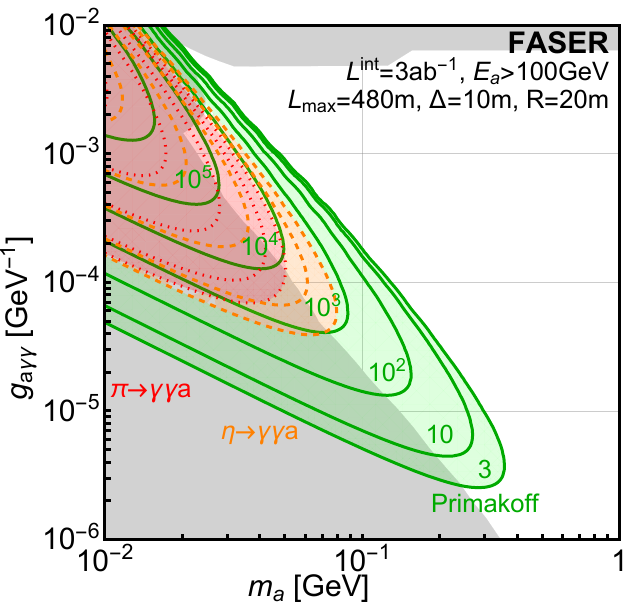} 
\includegraphics[width=0.48\textwidth]{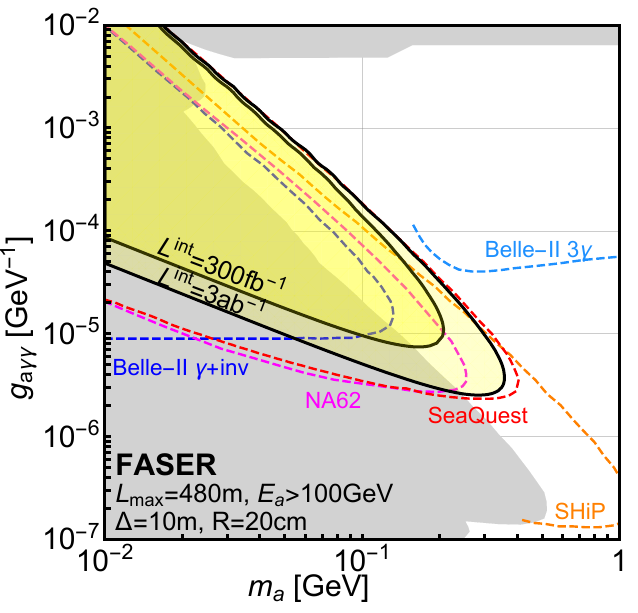} 
\caption{Left: For the ALP production mechanisms indicated, the number of ALP decays in FASER in the $(m_a, \ga)$ parameter space, given an integrated luminosity of $3~\iab$.  Right: Reach for FASER ($N=3$ signal events) for integrated luminosities $300~\ifb$ and $3~\iab$. For comparison we also show the projected reach of other proposed ALP searches.  The projected reach at Belle-II assumes the full expected integrated luminosity of $50~\iab$~\cite{Dolan:2017osp}. The reach for NA62 assumes $\sim 3.9 \times 10^{17}$ protons on target (POT) while running in a beam dump mode that is being considered for LHC Run 3~\cite{Dobrich:2015jyk}. The SeaQuest reach assumes $\sim 1.44 \times 10^{18}$ POT, which could be obtained in two years of parasitic data taking and requires additionally the installation of a calorimeter~\cite{Berlin:2018pwi}. The reach for proposed beam dump experiment SHiP assumes $\sim 2 \times 10^{20}$ POT collected in 5 years of operation~\cite{Dobrich:2015jyk}. }
\label{fig:yield}
\end{figure}

The ALP-induced signal in FASER typically consists of two high-energy photons coming from the ALP decay inside the detector. In the left panel of \figref{yield} we show the expected signal event yield in FASER in the $(m_a,\ga)$ plane, assuming an integrated luminosity of $3~\iab$. The gray shaded regions, which are adapted from Ref.~\cite{Dolan:2017osp}, represent the parameter space that is already excluded by previous experiments. The colored contour lines correspond to the number of ALP decays within FASER's decay volume, for the ALP production mechanisms indicated. As can be seen, the Primakoff process indeed provides the leading contribution, while meson decays only add a $\mathcal{O}(10\%)$ correction for most parts of parameter space. 

Notably, up to $\sim 10^5$ signal events can be expected in still unconstrained regions of parameter space. Note that FASER mainly probes the region of parameter space in which ALPs are required to be highly boosted to reach the detector. This is exactly the regime in which FASER has been shown to have significant discovery potential, comparable to the reach of SHiP for the case of dark photons~\cite{Feng:2017uoz}.  Note also that in the upper part of the region covered by FASER, the lines with constant number of signal events are very tightly spaced. In this regime, the decay length is significantly smaller than the distance to the detector, $\bar{d}_a\ll \lmin$, resulting in a strong exponential suppression of the number of events once the decay length drops further, as discussed in Ref.~\cite{Feng:2017uoz}. This limits the ability to probe higher $\ga$, but also implies that the reach in this region of parameter space is highly insensitive to the number of background events and to the signal detection efficiency. 

In the right panel of \figref{yield}, we show FASER's projected total sensitivity in the ALP parameter space. Here we assume that backgrounds can be reduced to a negligible level and a signal acceptance of $100\%$. A more detailed discussion is postponed until \secref{detection}. We note that even the subdominant decay channel $a \to e^+ e^- \gamma$, which only has a branching fraction of $\sim 1\%$, may also be able to cover unprobed parameter space. 

For comparison, we also show future projections of the sensitivity reach for Belle-II~\cite{Dolan:2017osp}, as well as for the beam dump experiments NA62~\cite{Dobrich:2015jyk}, SeaQuest~\cite{Berlin:2018pwi}, and SHiP~\cite{Dobrich:2015jyk}.  In the parameter space with $\ga \sim 3 \times 10^{-6} - 3 \times 10^{-2}$, FASER's reach is comparable to or better than the projected future sensitivities of these other experiments. As discussed above, in the regime of $\bar{d}_a\ll \lmin$, the contours with fixed number of signal events are very close to each other. As a result, the sensitivity reaches of FASER and the other experiments are similar, despite significant differences in luminosity. The effect of the increase in luminosity can, however, be observed at low values of coupling constants $\ga$ that allow larger lifetime. In this regime, ALPs can be less boosted and still reach FASER. However, such less energetic ALPs are typically characterized by larger $\theta_a$ and they miss FASER. This disadvantage is less pronounced for a much larger detector like SHiP.

\begin{figure}[t]
\centering
\includegraphics[width=0.48\textwidth]{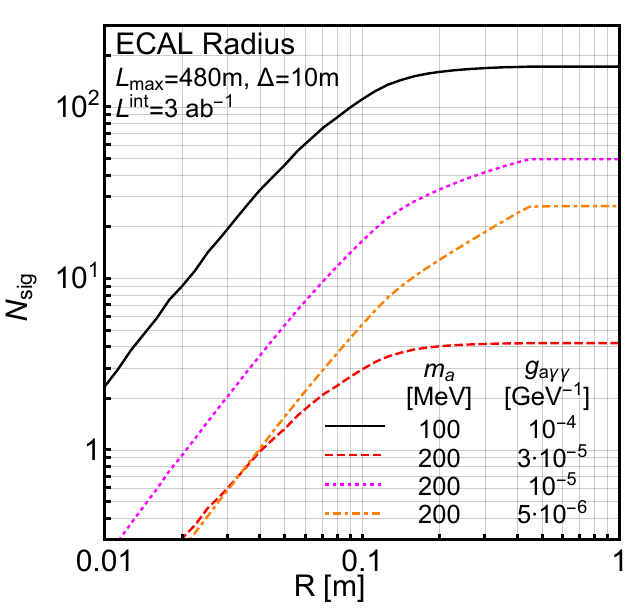} 
\includegraphics[width=0.48\textwidth]{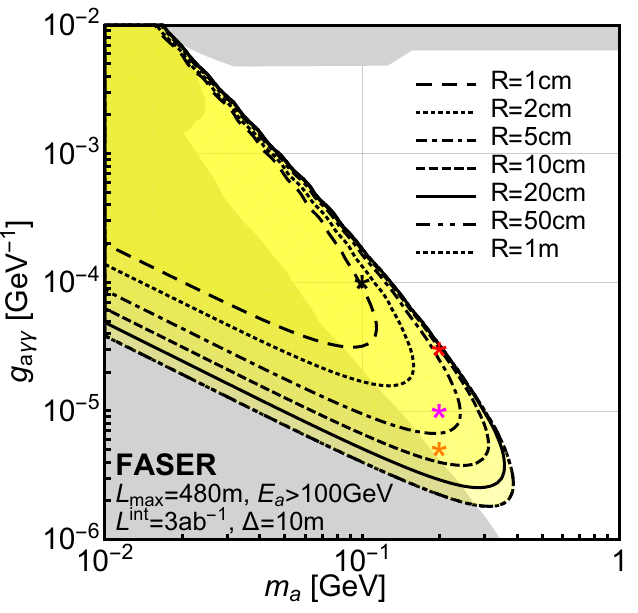} 
\caption{Left: Number of ALP decays as a function of radius $R$ for various choices of $(m_a, \ga)$. Right: Reach for FASER for $3~\iab$ in ALP parameter space for different values of the radius $R$. The stars corresponds to the benchmarks in the left panel. These results show that the ALP signal is highly collimated within distances $\sim {\cal O}(10)~\cm$ of the line of sight, and even a small detector with radius $R \sim 1~\cm$ may probe currently unconstrained regions of ALP parameter space.}
\label{fig:radius}
\end{figure}

In the left panel of \figref{radius}, we show the number of signal events as a function of FASER's radius $R$ for several benchmark values of $m_a$ and $\ga$.  In the right panel, the sensitivity reach in the $(m_a,\ga)$ plane is shown for several values of the radius $R$. As can be seen, even a very small detector with $R=2~\cm$ can probe unconstrained regions of ALP parameter space. Increasing $R$ above $\sim 10~\cm$ has a very mild impact on the reach for larger values of $\ga$.

\section{Detection of a Di-Photon Signal in FASER}
\label{sec:detection}

The ALP-induced signal in FASER typically consists of two highly-collimated, highly-energetic ($\sim \tev$) photons that point back to the IP. Given a detector consisting of several layers of tracker followed by an EM calorimeter, the ALP signal could be detected in the EM calorimeter or, if the photons convert into $e^+e^-$ pairs, in the tracking system. 

Given the shielding of the detector from the main LHC tunnel, one does not expect high-energy electromagnetic particles produced in beam-induced collisions in the beam pipe to reach FASER. Also, hadronic particles that could induce partly electromagnetic showers when interacting in the concrete or rock before FASER or inside the detector are expected to first effectively lose their energy. This suppresses the SM background for energies of the incoming ALP that are above a certain threshold. To determine the remaining background, a detailed FLUKA simulation for FASER is currently being performed, and there are plans to validate these simulations with in situ measurements. 

If the number of ALP-induced high-energy signal events significantly surpasses the expected number of background events, the detection of ALP events as a single high-energy EM shower without accompanying tracks might be sufficient to indicate the discovery of new physics.  On the other hand, if the background to such ``single photon" events is significant, distinguishing the two photons produced in ALP decays, that is, detecting the ALP signal as genuine di-photon events, may be required.  The background to high-energy di-photon events from the direction of the IP is essentially negligible.  
 
In the lab frame the distribution of the opening angle between two photons, $\theta_{\gamma\gamma}$, is strongly peaked towards its minimal value, $\theta_{\gamma\gamma}\simeq 2/\gamma$, where $\gamma=E_{a}/m_{a}$ is the ALP's boost factor. For typical ALP energies $E_a\sim \tev$ and $m_a\sim 100~\mev$, the opening angle is $\theta_{\gamma\gamma}\sim 200~\murad$. After 1 meter, this leads to a small separation between the two photons of order $d_{\gamma\gamma}\sim 200~\micm$, which makes it challenging to resolve them in a calorimeter. 

Remarkably, however, the required resolution might be achieved by employing existing calorimeter technology.  In particular, such resolutions are already achieved in the calorimeters used by the LHCf collaboration~\cite{Adriani:2008zz}. These consist of 16 layers of plastic scintillators interleaved with tungsten converters, which increase the radiation length of the detector and minimize the shower leakage effects. The total longitudinal size of such a calorimeter ``tower'' is about $23~\cm$, small enough to fit in the FASER location without sacrificing much tracker volume. The calorimeter's energy resolution is roughly $5\%$ for $E_\gamma> 100~\gev$~\cite{Mase:2012zz} and is employed to study photons originating from neutral pion decays with energies up to several $\tev$~\cite{Adriani:2015iwv}. The spatial resolution for the initial position of the photon entering the calorimeter can be better than $200~\micm$~\cite{Mase:2012zz}. This is achieved with four layers of microstrip silicon sensors that are placed within the calorimeter towers. Most important for the present context, events with two photons that develop two distinct peaks in the lateral shape of the shower can be distinguished with more than $90\%$ accuracy provided that the peaks are $\gtrsim 1~\mm$ apart from each other~\cite{ADRIANI:2013ira}. This can be done assuming that the lower energy photon carries at least $5\%$ of the energy of the more energetic one, which is almost always the case in ALP decay.

The possibility of distinguishing two nearby photons has also been studied for high-$p_T$ searches at ATLAS and CMS. For example, at ATLAS, for energy deposited in a calorimeter, a variable $w_{s3}$ is defined, which corresponds to the ratio of energy deposited in the two strips adjacent to the central one relative to the total energy deposited in all three strips. This has been used in Ref.~\cite{Draper:2012xt} to differentiate di-photon and single-photon events and translated into a limit on the difference in pseudorapidity, below which two photons are indistinguishable.  Such a limit corresponds to about a half of the strip size in the first layer of the electromagnetic calorimeter, which roughly leads to a spatial separation $\delta\sim 1-2~\mm$~\cite{CMS-PAS-EGM-10-006,Aaboud:2016yuq}. Other techniques involving more sophisticated photon-jet substructure analyses can also be used for a better discrimination~\cite{Toro:2012sv, Ellis:2012zp, Ellis:2012sd}.

\begin{figure}[t]
\centering
\includegraphics[width=0.48\textwidth]{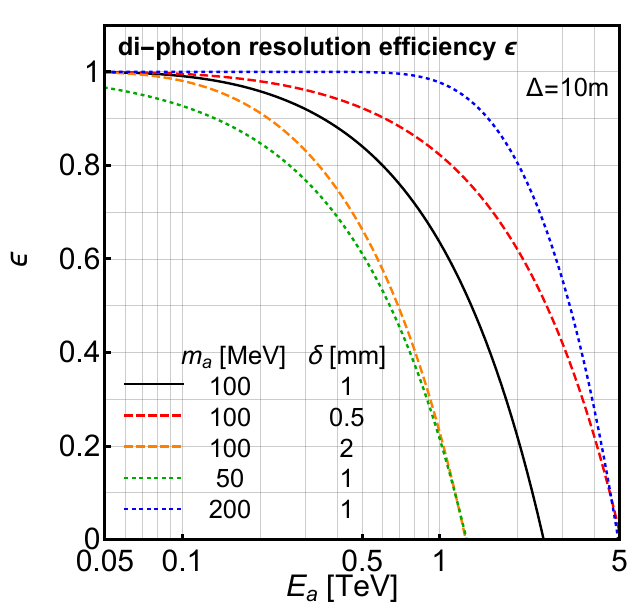} 
\includegraphics[width=0.48\textwidth]{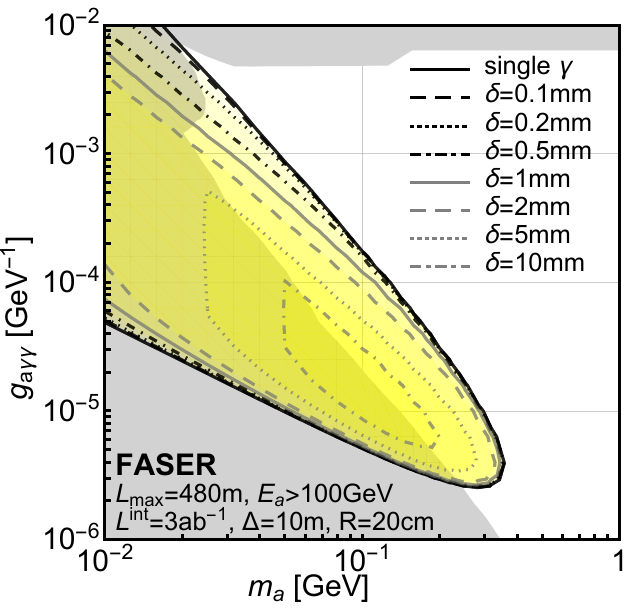} 
\caption{Left: Efficiency $\epsilon$ to detect the di-photon signal as a function of ALP energy for different choices of ALP mass $m_a$ and detector resolution $\delta$, the minimal separation of two photons that can be distinguished as two photons in the EM calorimeter.  The detector depth is set to $\Delta = 10~\m$; to an excellent approximation, $\epsilon$ depends on the ratio $\delta / \Delta$. Right: The reach in ALP parameter space for the di-photon signal for various values of $\delta$.}
\label{fig:twophoton}
\end{figure}

Requiring that the two photons decay products of the ALP are separated by a calorimeter spatial resolution $\delta$, so that they can be resolved as two photons, effectively reduces the depth of the detector.  
For the example above with $\theta_{\gamma\gamma}\sim 200~\murad$, requiring a spatial separation $\delta\sim 1~\mm$ typically requires that the photons travel a distance $\sim 5~\m$ in the detector. Of the ALPs that decay in the detector, then, the number that have photon separations greater than $\delta$ is effectively given by the number that decay in the reduced depth $\Delta_{\rm red} = \Delta - \delta/\theta_{\gamma\gamma}$.  Since the number of events depends on the depth as shown in \eqref{eq:decay_in_vol}, this reduces the number of signal events.  The efficiency for di-photon detection, that is, the fraction of ALP decays in the detector that can be resolved as di-photon events, given a detector resolution $\delta$, is
\be
\epsilon\equiv\frac{N_{\rm ev}^{\rm res}}{N_{\rm ev}} \sim \Bigg\langle\frac{1-\exp{\left(-\Delta/\bar{d}_a \right)}\exp{\left[ \delta/(\bar{d}_a \, \theta_{\gamma\gamma})\right]}}{1-\exp{\left( -\Delta/\bar{d}_a \right)}}\Bigg\rangle \approx 1-\frac{\delta}{\Delta}\times \frac{E_a}{2\,m_a} \ ,
\label{eq:efficiencycorrection}
\ee
where $\langle\cdots\rangle$ denotes the average over the distribution of opening angles $\theta_{\gamma\gamma}$.  In the last step, to provide a rough, but simple, approximation, we set the photon-photon opening angle to the fixed value $\theta_{\gamma\gamma} = 2m_a/E_a$, and assume $\Delta \ll \bar{d}_a $ and $\delta / \bar{d}_a \ll \theta_{\gamma\gamma}$.  The approximation is quite accurate when $\delta / \Delta \ll 2 m_a / E_a$ and the deviation of $\epsilon$ from 1 is small, but it breaks down for $\delta / \Delta \agt 2 m_a / E_a$, when the full simulation result must be used.  In our numerical results we employ the exact $\theta_{\gamma\gamma}$ distribution.  In the left panel of \figref{twophoton}, we show di-photon efficiencies $\epsilon$ as a function of the ALP energy for fixed $\Delta = 10~\m$ and some representative choices of ALP mass and detector resolution $\delta$. 

The precise value of $\delta$ will depend on the final detector setup and technology, but we note that for the aforementioned case with $\delta\simeq 1~\mm$ and for the detector dimensions given in \eqref{eq:far_location}, we obtain $\delta/\Delta = 10^{-4}$, and the typical suppression factor for detecting the di-photon signal from ALP decays in FASER is about 50\%. In the right panel of \figref{twophoton} we present the impact this can have on the sensitivity reach. The reach for several other choices of $\delta/\Delta$ are also shown for comparison. In particular, the case with $\delta/\Delta=0$ corresponds to the scenario with negligible background when even an effectively single-photon signal is enough for discovery.

If an ALP-like signal is observed in the calorimeter, further improvement of the analysis requires proper particle identification. In particular, FASER is also sensitive to many models for physics beyond the SM that lead to a signal that consists of high-energy electron-positron pairs. These could be disentangled from di-photon events by the use of a tracker and by employing a sufficiently strong magnetic field. On the other hand, hadronic neutral particles depositing their energy in calorimeters could be differentiated from photons based on their distinct shower development. Examples of such analyses are given in Refs.~\cite{Adriani:2015nwa,Adriani:2015iwv}, where hadronic neutral showers and EM showers are differentiated based on parameters $L_{20\%}$ and $L_{90\%}$, where $L_{n\%}$ denotes the length after which $n\%$ of the shower energy has been deposited in the calorimeter.

Photon conversion into $e^+e^-$ pairs inside the detector, in particular in the tracking system, can also allow one to disentangle single- and di-photon events; see, \eg, Ref.~\cite{Dasgupta:2016wxw} for a recent discussion. In the ATLAS detector such a conversion can occur in $10-50\%$ of the cases~\cite{Aad:2010sp}, depending on the pseudorapidity, making it an important search strategy. However, for FASER a dedicated analysis is needed to determine whether this approach can be used, given that the trackers will only constitute a small fraction of the total decay volume. In the case of conversion of one of the photons, one expects a signal in the calorimeter that consists of three simultaneous electromagnetic showers, one from the second photon and two from the $e^+$ and $e^-$ deflected by the magnetic field.

\section{Conclusions}
\label{sec:conclusions}

Searches for new light, weakly-coupled particles could provide the first evidence of physics beyond the standard model, with wide-ranging implications for particle physics and cosmology.  This possibility has stimulated a variety of proposals for experiments that could discover these new particles, and it motivates studies to determine the reach and promise of these proposed experiments.

In this study, we have considered FASER, the ForwArd Search ExpeRiment at the LHC.  Previous studies have shown that FASER can harness the currently ``wasted'' large forward cross section in $pp$ collisions at the LHC to search for new light particles with renormalizable couplings to the SM, such as dark photons, dark Higgs bosons, and heavy neutral leptons~\cite{Feng:2017uoz,Feng:2017vli,Kling:2018wct,Batell:2017kty,Helo:2018qej,Bauer:2018onh}.  Even a small $\sim 1~\m^3$ detector that takes data concurrently with the ongoing high-$p_T$ experiments can achieve world-leading sensitivities to these types of new particles.  

Here we have determined the reach of FASER to a qualitatively different form of new physics: ALPs, which couple dominantly through non-renormalizable di-photon interactions.  Such ALPs are dominantly produced not at the IP, but by TeV-energy photons from the IP that collide with the neutral particle absorber (TAN or TAXN) $\sim 130~\m$ downstream.  These interactions produce high-energy ALPs through the Primakoff process, and these ALPs propagate through matter without interacting and mainly decay to two photons in FASER.  This process exploits FASER's capability as a {\em high-energy photon beam dump experiment}.

Our results show that ALPs produced in this way are highly collimated.  At FASER's location 480 m from the IP, for most underlying model parameters, most of the ALP signal is contained within $\sim 10-20~\cm$ of the beam collision axis. In this way, the ALP signal is similar to the dark photon signal, and both are more collimated than the dark Higgs and HNL signals.  With a detector spanning this area and $\sim 3-10~\m$ deep, we have shown that FASER could detect as many as $\sim 10^5$ ALP events at the HL-LHC and have sensitivity comparable to or better than other proposed experiments.  

Of course, another important way in which ALPs differ from other dark sector candidates is that their signal is not two charged tracks, but two photons with $\sim \tev$ energies that originate from the direction of the IP in time with bunch crossings.  FASER's sensitivity therefore depends on its calorimeter capabilities and the relevant EM shower backgrounds.  If the background of $\sim \tev$ EM showers with the required direction and timing is negligible, all ALP decays may be taken as a background-free signal.  Alternatively, if the EM shower background is non-negligible, the ALP signal of two photons can still be background-free, provided the two photons can be differentiated from each other.  Because the photons are highly collimated, this requires a calorimeter that can differentiate showers separated by $\sim 1~\mm$.  Remarkably, calorimeters with resolution $\delta \sim 1~\mm$ already exist, as discussed in \secref{detection}.  We have shown how the ALP reach depends on $\delta$.  With the existing technology, the efficiency for detecting di-photon signals can still be as large as $\sim 50\%$, and the reach in ALP parameter space is degraded only slightly.  Further progress depends on background simulations and in situ measurements that are currently underway.

In this work we have considered the case of axion-like particles coupling to photons, with both the coupling and mass as free parameters of the model. Within this framework, probably the most motivated model is the QCD axion, for which the coupling to photons is $\ga = c_a \alpha_{\text{EM}} m_a / ( 2 \pi m_\pi f_\pi)$, where the range for the prefactor $c_a$ is typically taken to encompass the values $\sim -4$ (KSVZ) to $\sim 1.5$ (DFSZ)~\cite{Agashe:2014kda}. For a QCD axion with mass $m_a=100~\mev$, this implies $\ga \sim 0.01$, which is excluded by beam dump experiments. However, recent work~\cite{Alves:2017avw} has shown that it is possible to construct a viable QCD axion at the MeV scale by coupling it to first-generation fermions, while keeping its mixing with the neutral pion suppressed. The di-photon coupling of such an axion can also be kept below current bounds. The possibility of ALPs with dominantly di-photon couplings but also other couplings is quite general~\cite{Jaeckel:2010ni}, and it is interesting to note that these models can also be probed by FASER in the way described here.

Finally, it is important to note that ALPs may also couple dominantly to other SM particles, such as gluons or fermions, and these couplings in fact induce each other at the loop level.  These alternative couplings may alter the di-photon signal and rate, allow ALPs to be produced through other processes, such as the rare decays of heavier mesons, or induce other ALP signals in FASER, such as the two charged track signals already analyzed for other dark sector candidates.

\acknowledgments

We thank Jamie Boyd, Dave Casper, Francesco Cerutti, Shih-Chieh Hsu, Simon Knapen, Mike Lamont, Hiroaki Menjo, Brian Shuve, and Yotam Soreq for useful discussions.  This work is supported in part by NSF Grant No.~PHY-1620638.  J.L.F. is supported in part by Simons Investigator Award \#376204.  I.G. is supported in part by DOE grants DE-SC0013678 and DOE-SC0010008.  S.T. is supported in part by the Polish Ministry of Science and Higher Education under research grant 1309/MOB/IV/2015/0 and by the National Science Centre (NCN) research Grant No.~2015-18-A-ST2-00748.

\appendix

\section{Kinematics of the Primakoff Process}
\label{sec:Primakoffkinematics}

In the Primakoff process, a photon converts to an ALP when passing through the electromagnetic field of an atom or nucleus with mass $M$. Given typically low momentum transfers between the projectile photon and the target nucleus, we can effectively describe this process as an elastic scattering of the photon on the target, $ \gamma N \to a N$.  Let us consider this process in the target's rest frame, where the initial state photon and target momenta are $p_\gamma^\mu = (E_\gamma, 0, 0, E_\gamma)$ and $p_i = (M,0,0,0)$, and parametrize the outgoing ALP momentum as $p_a^\mu = (E_a, p_a\sin\theta_{a\gamma}, 0, p_a\cos\theta_{a\gamma})$, where $\theta_{a\gamma}$ is the angle between the incoming photon and the ALP. The ALP momentum $p_a$ can be expressed in terms of the scattering angle $\theta_{a\gamma}$ as
\be
 p_a \!  = \!  \frac{ E_\gamma ( E_\gamma M \! +\! \frac{m_a^2}{2}) \cos\theta_{a\gamma}+(E_\gamma\!+\!M)
\sqrt{ \frac{m_a^4}{4}\!+\! E_\gamma^2 M^2 \!-\! m_a^2 \left[M (E_\gamma \!+\! M) + E_\gamma^2 \sin^2 \! \theta_{a\gamma}\right] }
}{ E_\gamma^2 \sin^2 \! \theta_{a\gamma} +2 E_\gamma M +M^2 }  .
\ee

For FASER we are mainly interested in high energy photons and small momentum transfers, which correspond to small-angle ALP scatterings.  For $m_a/\sqrt{M(M+2E_\gamma)},~\theta_{a\gamma} \ll 1$, we can approximate the ALP momentum as
\be
p_a = E_\gamma -\frac{m_a^2}{2E_\gamma} -\theta_{a\gamma}^2 \frac{E_\gamma}{2M} \left(E_\gamma-\frac{m_a^2}{2M}\right) \ .
\ee
The momentum transfer between the photon and target is
\be
t= -q^2 = -(p_a^\mu - p_\gamma^\mu)^2 = 2E_\gamma(E_a - p_a \cos\theta_{a\gamma}) - m_a^2
\approx \frac{m_a^4}{2E_\gamma^2}+E_\gamma^2\theta_{a\gamma}^2 \ .
\ee

\begin{figure}[t]
\centering
\includegraphics[width=0.32\textwidth]{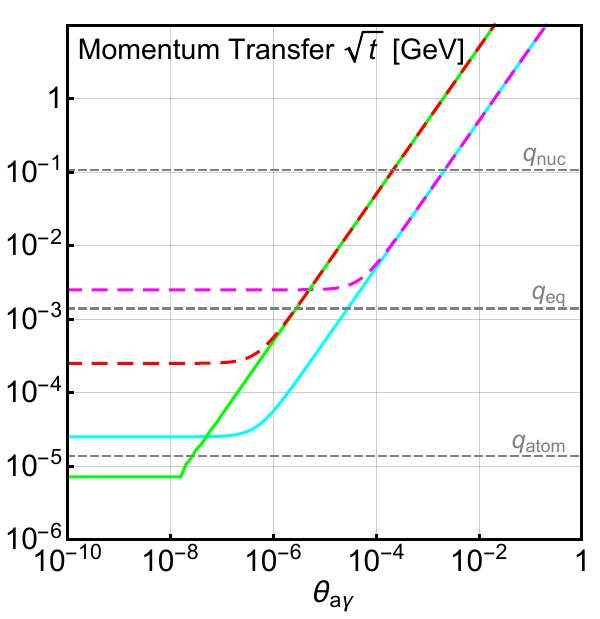}
\includegraphics[width=0.32\textwidth]{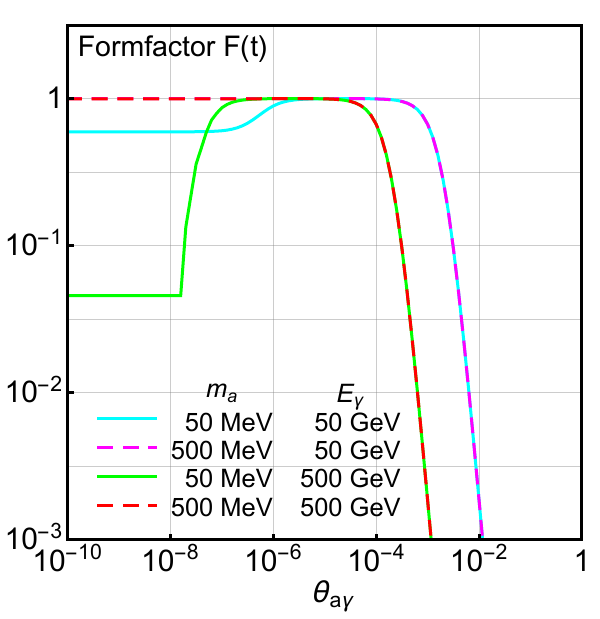}
\includegraphics[width=0.32\textwidth]{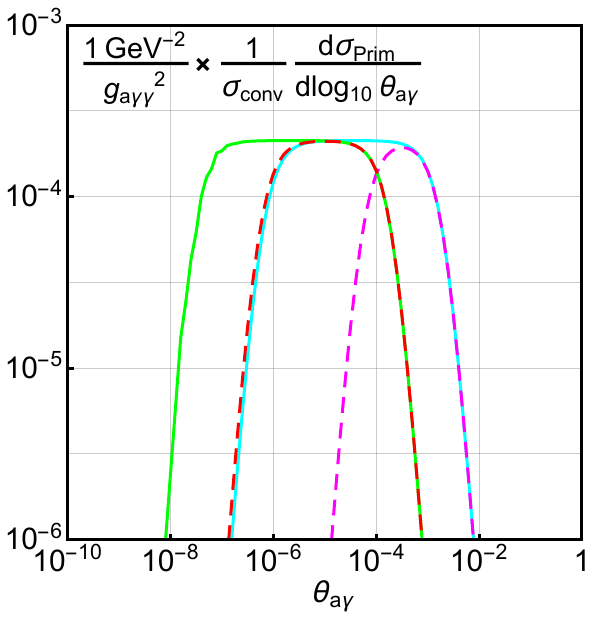}
\caption{Momentum transfer $\sqrt{t}$ (left), form factor $F(t)$ (center), and the Primakoff differential cross section of \eqref{eq:cross_sec_alp_prod} (right) for different combinations of ALP mass $m_a$ and photon energy $E_\gamma$. These results are for scattering off iron with $Z_{\rm nuc}=26$ and $A_{\rm nuc}=56$.
 }
\label{fig:Primakoff_dists}
\end{figure}

The left panel of \figref{Primakoff_dists} shows the momentum transfer $q=\sqrt{t}$ as a function of the scattering angle for various ALP masses and photon energies.   At angles much smaller than $\theta^* \equiv m_a^2/(\sqrt2 E_\gamma^2)$, $t$ approaches a constant value $t_{\text{min}} = m_a^4/(2E_\gamma^2)$, while, for larger angles, it scales as $t \approx E_\gamma^2 \theta_{a\gamma}^2 $.  The horizontal gray dashed lines show typical scales of the momentum transfer for the form factors defined in \eqref{eq:form_factors}: the atomic form factor scale $q_{\text{atom}}=1/a= Z_{\text{nuc}}^{1/3} m_e / 111$, the atomic-nuclear form factor crossover scale $q_{eq}=2.71 \, m_e$, and the nuclear form factor scale $q_{\text{nuc}}=\sqrt{d}=0.4 A_{\text{nuc}}^{-1/3}~\gev$. Note that depending on $m_a$ and $p_a$, the atomic form factor and its cutoff scale might or might not be relevant for ALP production. 

The form factor $F(t)$ is shown in the central panel of \figref{Primakoff_dists}. We see that (1) for $|q| \gg q_{\text{nuc}}$, the form factor decreases as $F(q) \sim q_{\text{nuc}}^2 / q^2$, thus suppressing large angle $\theta_{a\gamma}$ (large momentum transfer) scattering; (2) for $q_{\text{atom}} \ll |q| \ll q_{\text{nuc}}$, the form factor is approximately equal to unity; and (3) if $|q|<q_{\text{atom}}$ is kinematically accessible, the form factor approaches $F(t_{\text{min}})$ as $\theta_{a\gamma}$ decreases.

For scattering angles $\theta_{a\gamma} \gg \theta^*$ the momentum transfer scales like $t \approx E_\gamma^2 \theta_{a\gamma}^2 $, and we can approximate the differential Primakoff cross section in \eqref{eq:cross_sec_alp_prod} as\footnote{
For $\theta^* < \theta_{a\gamma} \ll 1$,
\be
\theta_{a\gamma}\frac{p_a^4 \sin^3\theta_{a\gamma} }{t^2}
\approx
1- \frac{m_a^2}{E_\gamma^2}
\left( 2 + \frac{m_a^2}{E_\gamma^2\theta_{a\gamma}^2} \right)
- \frac{2}{M}
\left( \theta_{a\gamma}^2-\frac{m_a^4}{E_\gamma^4} \right)
\left( E_\gamma - \frac{m_a^2}{2M} \right)
\nn
\ee
exhibits a plateau structure: a steep ascent, a constant region, and a steep descent. At larger angles it asymptotes to $\approx 1$.
}    
\be
\frac{d\sigma_{\text{Prim}}}{d\log \theta_{a\gamma}} \bigg|_{\theta_{a\gamma} \gg \theta^*} 
=  \theta_{a\gamma} \frac{d\sigma_{\text{Prim}}}{d\theta_{a\gamma}} 
\approx \frac{1}{4} \ga^2 \alpha Z^2 
 F^2(t) \, .
\ee
This accounts for the steep descent at large angles for which $t \gtrsim q_{\text{nuc}}$, and the constant plateau region, where the function $F^2(t)$ is approximately constant.  As $\theta_{a\gamma}$ approaches $\theta^*$, $t$ decreases, and the form factor approaches a constant value $F(t_{\text{min}})$.  At this kinematic region the differential cross section is given by
\be
\frac{d\sigma_{\text{Prim}}}{d\log\theta_{a\gamma}} \bigg|_{\theta_{a\gamma} < \theta^*} 
=  \theta_{a\gamma} \frac{d\sigma_{\text{Prim}}}{d\theta_{a\gamma}} 
\approx \frac{\ga^2 \alpha Z^2}{4}\frac{ F^2(t_{\text{min}}) E_\gamma^4}{t_{\text{min}}^2}\, \theta_{a\gamma}^4 \ ,
\ee
which decreases as $\sim \theta_{a\gamma}^4$.  These results, normalized to the conversion cross section of photons in iron, are shown in the right panel of \figref{Primakoff_dists}.  Note that the transition at $\theta_{a\gamma}=\theta^*$ depends on the ratio $m_a/E_\gamma$. For FASER energies and masses of interest, this transition typically occurs when the form factor is above the atomic scale cutoff $q_{\text{atom}}$.

The total cross section for photon conversion into an ALP via the Primakoff process relative to the photon conversion cross section into electrons is shown in the left panel of \figref{production-rate} for different ALP masses and energies. For large $E_\gamma/m_a$, the Primakoff cross section approaches a constant maximum. In this case, a fraction of $\mathcal{O}(0.1\%) \times [\ga/\gev^{-1}]^2$ of the photons convert into ALPs. 

In summary, we have seen that the momentum transfer in the Primakoff process is typically small due to a cutoff of the nuclear form factor at $t >q_{\text{nuc}}^2$. For high energy photons, this implies small scattering angles of the ALP with respect to the photon direction, $\theta_{a\gamma} < q_{\text{nuc}}/E_\gamma$. Therefore, the ALP momenta are almost collinear with the photon momenta. Furthermore, the ALPs carry almost the entire energy of the photon $E_{a} \approx E_\gamma$. Hence the collinear approximation, $p_a \approx p_\gamma$, gives an excellent estimate for the final sensitivity reach discussed in the main text.

\section{Production of Axion-like Particles in Pion Decay}
\label{sec:piondecay}

\begin{figure}
\includegraphics[width=0.48\textwidth]{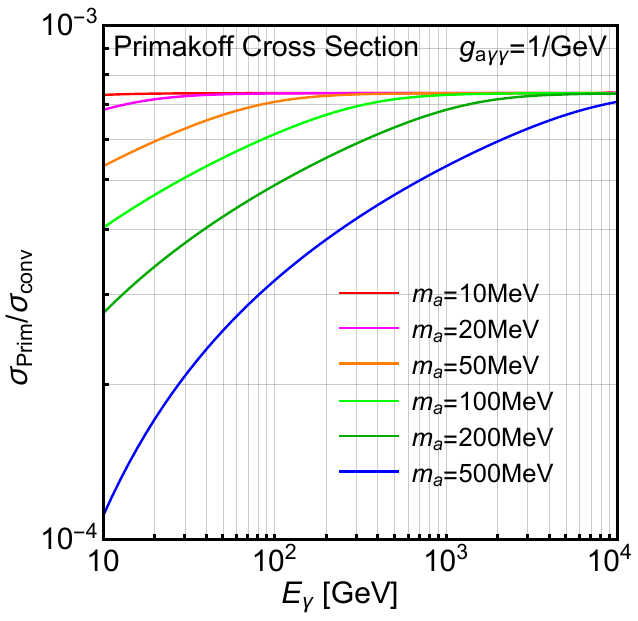} 
\includegraphics[width=0.48\textwidth]{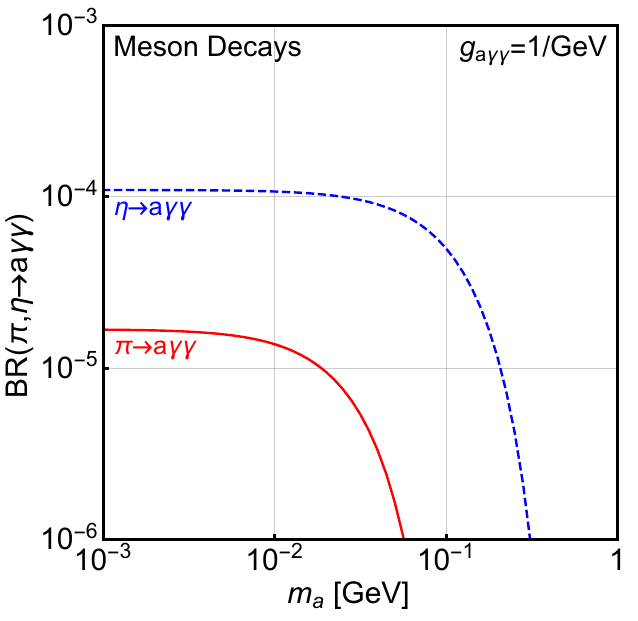}
\caption{Left: The ratio between the Primakoff and pair-production cross sections, $\sigma_{\text{Prim}}/\sigma_{\text{conv}}$, as a function of the photon energy $E_\gamma$, for different values of $m_a$ and $\ga=1~\gev^{-1}$. Right: The branching ratios $B(\pi^0, \eta \to a\gamma\gamma)$ as functions of $m_a$ for $\ga = 1~\gev^{-1}$.} 
\label{fig:production-rate}
\end{figure}

If the ALP is light enough, it can also be produced in the decays of neutral pions $\pi^0$. To calculate the decay width $\Gamma(\pi^0\to a \gamma\gamma)$, let us consider the interaction Lagrangian in the effective theory
\be
{\cal L} \supset
-\frac14\gpi \pi^0 F^{\mu\nu}\widetilde F_{\mu\nu}
-\frac14\ga a F^{\mu\nu}\widetilde F_{\mu\nu} \ ,
\ee
where $\gpi =2.512 \times 10^{-2}~\gev^{-1}$ is the pion decay constant and we conform to the convention that $\widetilde{F}_{\mu\nu} = \frac{1}{2} \epsilon_{\alpha\beta\mu\nu} F^{\alpha\beta}$ is the dual field strength tensor. Choosing a momentum assignment  $\pi^0(p)\to a(q)\gamma(k_1)\gamma(k_2)$, the decay amplitude is
\be
{\cal M} =-\gpi\ga\,\epsilon^{\alpha\beta\gamma\delta} \epsilon^{\mu\nu\rho\sigma} p_\alpha q_\mu g_{\beta\nu} 
\bigg[ \frac{k_{1\gamma}\varepsilon_{1\delta}k_{2\rho}\varepsilon_{2\sigma}}{(p-k_1)^2}
        +\frac{k_{2\gamma}\varepsilon_{2\delta}k_{1\rho}\varepsilon_{1\sigma}}{(p-k_2)^2} \bigg] \ .
\ee
Let us consider this process in the pion's rest frame, where the particle momenta are $p=(M,0,0,0)$, $k_1 = E_1(1,0,0,1)$ and $k_2 =E_2(1,\sin\theta_{12},0,\cos\theta_{12})$. Here $M$ denotes the pion mass. The ALP momentum is given by  $q = p-k_1-k_2$ with $q^2=m^2$, where $m$ is the ALP mass. Energy-momentum conservation implies 
\be
\cos\theta_{12} =  \frac{1}{2E_1E_2} \left[ M^2 - m^2 - 2M(E_1+E_2) + 2E_1E_2 \right] .
\ee

The differential decay width is, then,
\be
\frac{d\Gamma(\pi^0 \to a \gamma\gamma)}{dE_1 dE_2} 
= \frac{1}{2} \frac{1}{(4 \pi)^3 M}  \sum_{\rm{pols}} |M|^2 
= \frac{(\gpi \, \ga)^2}{2 (4\pi)^3 M}  f(E_1,E_2) \ ,
\ee
where
\be
f(E_1,E_2)
&= \frac{E_1^2 E_2^2 \big[ M^2 (1+\cos^2\theta_{12}) +2E_1 (E_1 - M)(1-\cos\theta_{12})^2 \big]}
{(M-2E_1)^2}\\
&+2 \frac{E_1^2E_2^2 \big[M^2(1+\cos^2\theta_{12})+[2E_1 E_2 - M (E_1+E_2)](1-\cos\theta_{12})^2\big]}
{(M-2E_1)(M-2E_2)}\\
&+
\frac{E_1^2 E_2^2 \big[ M^2 (1+\cos^2\theta_{12}) +2E_2 (E_2 - M)(1-\cos\theta_{12})^2 \big]}
{(M-2E_2)^2} \ .
\ee
Integrating over phase space results in the total decay width
\be
\Gamma(\pi^0\to a \gamma\gamma) 
&= \! \!
\int_0^{\frac{M^2-m^2}{2M}} \!\!\!\!\!\! dE_1
\int_{\frac{M^2-m^2}{2M} - E_1}^{\frac{M}{2}+ \frac{m^2}{4E_1-2M}} \!\!\! dE_2
\frac{d\Gamma(\pi^0 \to a \gamma\gamma)}{dE_1 dE_2}
= \frac{(\gpi \ga)^2}{768 (4\pi)^3  M^3} \,  F(M,m) \ ,
\ee
where
\be
\!\!\!\! F(&M,m)= 24 \log \left(\!\frac{m}{M}\!\right)\! 
\left[6 m^2 M^2(M^4\!+\!m^4)\!+\!15m^4 M^4 \!+\! 2 m^4 M^4\log \left(\!\frac{m M}{m^2\!+\!M^2}\!\right)\right]
\\
&+7 (M^8\!-\!m^8) \!+\!148M^2 m^2(M^4\!-\!m^4) \!+\!24 m^4 M^4
\left[
\text{Li}_2\left(\!\frac{m^2}{m^2\!+\!M^2}\!\right)\!-\!
\text{Li}_2\left(\!\frac{M^2}{m^2\!+\!M^2}\!\right)\!
\right].
\ee
Similar results can be obtained for the $\eta$ meson, where one can use $\gpi\approx \geta$ (up to an ${\cal O}(10^{-4})$ correction). The branching fractions for both $\pi^0$ and $\eta$ decays as functions of the ALP mass are given in \figref{production-rate} for $\ga = 1~\gev^{-1}$. As one can see, the ALP production rate in rare $\pi^0$ and $\eta$ decays is typically suppressed compared to the Primakoff process. (Cf. the right panel of \figref{alp_production}.) Note also that ALPs from 3-body meson decays are typically less boosted than ALPs produced in the Primakoff process.  As a result, meson decays are less significant for FASER event rates throughout parameter space, as can be seen in the left panel of \figref{yield}.

\section{The Angular Acceptance Function}
\label{sec:angularacceptance}

\begin{figure}[t]
\centering
\includegraphics[width=0.8\textwidth]{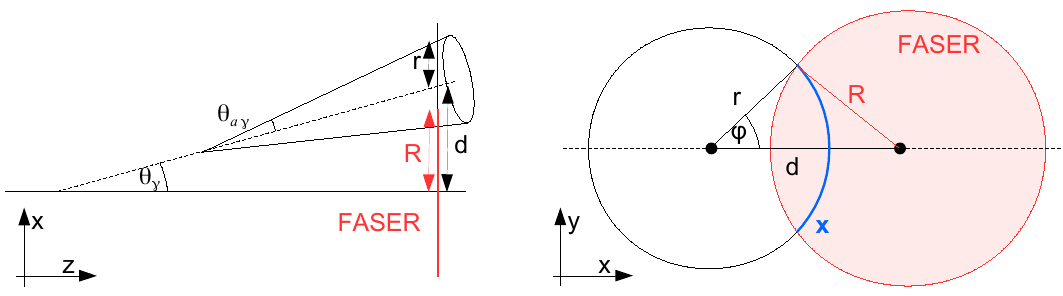} 
\caption{Left: Geometric setup for ALP production and detection in FASER. Right: Intersection of the cone for ALP production with fixed $\theta_{a\gamma}$ with the plane at $z=\lmax$ containing FASER.}
\label{fig:geometry}
\end{figure}

The angular distribution of ALPs produced at the TAXN in the Primakoff process can be inferred from \eqref{eq:cross_sec_alp_prod}. In particular, while the polar angle with respect to the photon direction $\theta_{a\gamma}$ is peaked at small values $\theta_{a\gamma}\approx 0$, the azimuthal angle remains uniformly distributed. This implies that ALP momenta are located in a cone around the photon propagation axis, as shown in the left panel of \figref{geometry}. Both the photon angle with respect to the beam collision axis $\theta_\gamma$ and the ALP scattering angles $\theta_{a\gamma}$ are characteristically small for the forward high-energy photons relevant for FASER, implying that the conic intersection at $z=\lmax$ is, to a good approximation, a circle with radius $r = \theta_{a\gamma}\,(\lmax-L_{\text{\TAN}})$ whose center is displaced from the beam collision axis by a distance $d = \theta_\gamma\,\lmax$. 

If FASER's geometry is cylindrical, FASER's intersection with this plane is a circle of radius $R$ whose center lies on the beam collision axis. FASER's angular acceptance function $\aang(\theta_\gamma, \theta_{a\gamma})$ for detecting such an ALP therefore has a simple geometric interpretation as shown in the right panel of \figref{geometry}: it is the ratio of the arc-length from the ALP circle overlapping with FASER, to the circumference of the ALP circle.  The angular acceptance can therefore be written as $\aang=1$ for $d+r<R$,  $\aang=0$ for $d-r>R$, and
\be
\aang = \frac{1}{\pi r} \int_0^{\varphi} \, r \,d\phi
= \frac{\varphi}{\pi}
= \frac{1}{\pi}\,  \cos^{-1} \left( \frac{d^2 + r^2- R^2}{2rd} \right)
\ee 
otherwise.  

\bibliography{alps_faser}

\end{document}